\documentclass[aps,prd,showpacs]{revtex4}
\usepackage{graphicx}% Include figure files
\usepackage{dcolumn}% Align table columns on decimal point
\usepackage{bm}% bold math
\usepackage{array,tabularx}      %all table
\usepackage{amsmath}%edit the mathmatical equation
\usepackage{mathrsfs} %\mathscr{AB}%
\usepackage{indentfirst}
\usepackage{slashed}
\usepackage{indentfirst}
\usepackage{slashbox}
\usepackage{multirow}
\usepackage{color}

\begin{document}
%\preprint{}
\title{Radiative vertex $VP\gamma$ and $\eta-\eta'$ mixing in light-cone sum rules}% Force line breaks with \\
\author{Jian-Mei Hou, Chang-Qiao Du, Hong Chen, Ming-Zhen Zhou}
\email{zhoumz@mail.ihep.ac.cn}
\affiliation{School of Physical
Science and Technology, Southwest University, Chongqing 400715,P. R.
China.}
%\author{Tao Huang}
%\affiliation{ CCAST (World Laboratory), P. O. Box
%8730,Beijing,100080
%\\Institute of High Energy Physics, P. O. Box 918, Beijing, 100039, P.
%R. China}%Lines break automatically or can be forced with \\

\date{\today}% It is always \today, today,

\begin{abstract}
In this work, we calculate radiative vertex $VP\gamma$
($V=\phi,\omega,\rho$ and $P=\eta,\eta'$) by utilizing $\omega-\phi$
mixing scheme and taking into account the contributions of the
three-particle twist-4 distribution amplitudes of the photon in QCD
sum rules on light cone. According to experimental data of
$V\rightarrow P\gamma$ and $P\rightarrow V\gamma$ from PDG, a value
of the $\eta-\eta^\prime$ mixing angle,
$\varphi=(40.9\pm0.5)^\circ$, is extracted in the framework of the
quark-flavor basis to describe the $\eta-\eta^\prime$ system.
\end{abstract}

\maketitle

\section{Introduction}

The problem of $\eta-\eta^\prime$ mixing in the pseudoscalar-meson
nonet has been studied many times in the last forty years. Because
those researches play a important role for the $SU(3)$-breaking
effect and the $U(1)_A$ anomaly. In contrast to the $\phi$ and
$\omega$ mesons, where they are taken as almost ideally mixed states
with quark content of well defined flavor, $\eta-\eta^\prime$ mixing
is still a debated subject. In the pioneering
work\cite{pioneer1,pioneer2,pioneer3}, a mixing angle $\theta_P$ was
conventionally introduced to describe $\eta$ and $\eta^\prime$ as
linear combinations of octet and singlet basis states. With the
development of the experiment, a phenomenological investigation
\cite{phinvest} found that one mixing angel was insufficient to
describe more physical processes, where $\eta$ or $\eta^\prime$
meson appears in the initial state or the final state. And then two
new equivalent schemes to describe the $\eta-\eta^\prime$ mixing was
proposed by Leutwyler\cite{oss} and Feldmann et al.\cite{qfs},
respectively. The correspondence mixing angels are
$\theta_1,\theta_8$ for the octet-singlet basis and
$\varphi_q,\varphi_s$ for the quark-flavor basis. More literatures
about two mixing angles can be found in
Refs.\cite{ta1,ta2,ta3,ta4,ta5}, where $\theta_1$ is not equal
$\theta_8$, but $\varphi_q$ is equal $\varphi_s$ apart from terms
which violate the Okubo-Zweig-Iizuka (OZI) rule\cite{voozi}. In our
next calculation, we chose $\varphi_q=\varphi_s=\varphi$ in the
quark-flavor scheme to investigate radiative vertex $VP\gamma$
($V=\phi,\omega,\rho$ and $P=\eta,\eta'$).

The radiative decays between light pseudoscalar (P) and light vector
(V) mesons are an excellent laboratory for investigating the nature
and extracting the non-perturbative parameters of light pseudoscalar
nonet in low-energy hadron physics. Among the characteristics of the
electromagnetic interaction processes, the coupling constant,
$g_{VP\gamma}$, plays one of the most important roles, since they
determine the strength of the hadron interactions. So by
investigating the above six radiative vertexes, in which $\eta$ or
$\eta^\prime$ meson was involved, one can extracting the mixing
angle of $\eta-\eta^\prime$ system. Interest on this issue has been
performed by many authors, for example. in
Refs.\cite{qfs,ta2,ravtx0,ravtx1,ravtx2,ravtx3}. In this paper, we
renewed this issue in QCD sum rules on light cone. Since the
radiative decays of light meson is belong to the low-energy hadron
interaction, which is governed by non-perturbative QCD, it is very
difficult to obtain the numerical values of the coupling constants
from the first principles. In order to interpret coupling constants
from the experimental data, we immediately need to deal with large
distance effects from the photon besides the hadrons, because a
special feature of the QCD description of hard exclusive processes
involving photon emission is that a real photon contains both a hard
electromagnetic and a soft hadronic component. In
Ref.\cite{photonDA1,photonDA2}, a consistent technique was proposed
by closed analogy with distribution amplitudes (DAs) of mesons
\cite{mesonDA1,mesonDA2}. The soft hadronic components of the photon
are related to matrix elements of light-cone operators with
different twist in the electromagnetic background field and can be
parameterized in terms of photon DAs. Since the photon emission from
the light quark takes place at large distances, the use of standard
QCD sum rules based on the local operator product expansion (OPE)is
not sufficient. Rather, one should use a light-cone expansion which
is adequate for exclusive processes with light particles.

The paper is organized as follows. In section 2, we derive the
coupling constants $g_{VP\gamma}$ by utilizing $\omega-\phi$ mixing
scheme and taking into account the contributions of the
three-particle twist-4 distribution amplitudes of the photon in the
light-cone sum rules. In section 3, we present our numerical
analysis. The final section is reserved for a conclusion. The photon
distribution amplitudes are list in Appendix A and the overlap
amplitudes for pseudoscalar mesons and vector mesons, which were
defined in section 2, are presented in Appendix B.

\section{ Radiative vertex $VP\gamma$ in light-cone sum rules }

In the framework of light cone QCD sum rules, we immediately choose
the two point correlation function with the photon as follow
\begin{equation}\label{hjm1}
T_{\mu}(p,q)=i\int
d^{4}xe^{-iqx}<\gamma(p)|T\{j_{\mu}^{V}(x)j_5^{P}(0)\}|0>,
\end{equation}
to extract the radiative coupling constant $g_{VP\gamma}$. Here
$j_\mu^V(x)$ is the vector meson current and $j_5^P(0)$ is the
pseudoscalar current. According to the quark model of hadron and
neglecting the contribution from the high Fock state of hadron, the
above interpolating currents may be written as
\begin{eqnarray}
j_{\mu}^{\phi}=
\frac{1}{\sqrt{2}}(\bar{u}\gamma_{\mu}u+\bar{d}\gamma_{\mu}d)\sin\beta+
\bar{s}\gamma_{\mu}s\cos\beta,~~
 j_\mu^{\omega}=
\frac{1}{\sqrt{2}}(\bar{u}\gamma_{\mu}u+\bar{d}\gamma_{\mu}d)\cos\beta-
\bar{s}\gamma_{\mu}s\sin\beta,~~
  j_\mu^{\rho}=
\frac{1}{\sqrt{2}}(\bar{u}\gamma_{\mu}u-\bar{d}\gamma_{\mu}d)
\end{eqnarray}
for light vector mesons $\phi,\omega,\rho$ and
\begin{eqnarray}
 j_5^{\eta}=
\frac{1}{\sqrt{2}}(\bar{u}i\gamma_{5}u+\bar{d}i\gamma_{5}d)\cos\varphi-
\bar{s}i\gamma_{5}s\sin\varphi,~~
  j_5^{\eta'}=
\frac{1}{\sqrt{2}}(\bar{u}i\gamma_{5}u+\bar{d}i\gamma_{5}d)\sin\varphi+
\bar{s}i\gamma_{5}s\cos\varphi
\end{eqnarray}
for light pseudoscalar mesons $\eta$,$\eta^\prime$, respectively.
Here $\varphi$ is the value of the $\eta-\eta^\prime$ mixing angle,
which will be discussed in this work, and $\beta$ is the value of
the $\omega-\phi$ mixing angle, which has been determined from the
available experimental data in the Ref.\cite{mixvector} as
$\beta=3.18^\circ$.

According to the basic assumption of quark-hadron duality in the QCD
sum rules approach, we can insert two complete series of
intermediate states with the same quantum numbers into the
correlation function $T_{\mu}$ to obtain the hadronic
representation. After isolating the contribution of the ground state
by the pole terms of the vector meson and the pseudoscalar meson, we
get the following result
\begin{equation}
T_{\mu}(p,q)=\frac{<P(-q)\gamma(p)|V(p-q)><V(p-q)|j_{\mu}^{V}|0><0|j_5^{P}|P(-q)>}{((p-q)^2-m^2_{V})((-q)^2-m^2_P)}+\ldots,
\end{equation}
for $V\rightarrow P\gamma$ decay, where $-q$, $p$ and $p-q$ denote
the pseudoscalar meson, the photon and the vector meson momentum,
respectively. The amplitudes of these interpolating currents with
the meson states are defined as
\begin{eqnarray}
<0|j_{\mu}^{V}|V(p-q)>&=&\lambda_{V}u_{\mu}^{V}\\
<0|j_5^{P}|P(-q)>&=&\!\!\!\lambda_{P},
\end{eqnarray}
where $u_{\mu}^{V}$ is the polarization vector of the vector meson,
$\lambda_V$ and $\lambda_P$ are called the overlap amplitudes which
can be determined by QCD sum rules method in Appendix B. The
coupling constant $g_{V P \gamma}$ is defined through the effective
Lagrangian
\begin{equation}
\mathscr{L}=-\frac{e}{m_{V}}g_{V P
\gamma}\varepsilon_{\mu\nu\alpha\beta}(\partial^{\mu}\phi^{\nu}_V-\partial^{\nu}\phi^{\mu}_V)(\partial^{\alpha}
A^{\beta}-\partial^{\beta}A^{\alpha})\phi_P
\end{equation}\\
where $\phi_V$,$\phi_P$ and $A$ denote the vector field, the
pseudoscalar field and the photon field, respectively. Therefore,
the $<P(-q)\gamma(p)|V(p-q)>$ matrix in the hadronic representation
can be written as
\begin{equation}
 <P(-q)\gamma(p)|V(p-q)>=-\frac{e}{m_{V}}g_{V P
\gamma}K(p^{2})\varepsilon^{\mu\nu\alpha\beta}u^V_\mu q^\nu
\varepsilon^\gamma_\alpha p^\beta ,
\end{equation}
where $p^{2}=0$ for the momentum of the real photon and $K(p^2)$ is
a form factor with $K(0)=1$. Substituting eq.(5),(6) and (8) into
the hadronic representation, we obtain the physical part and choose
the structure $\varepsilon_{\mu}^{\nu\alpha\beta}p_\nu
q_\alpha\varepsilon^\gamma_\beta$ from which the corresponding
invariant amplitude,
\begin{eqnarray}
T{((p-q)^2,q^2)}=\frac{e\ g_{V P
\gamma}\lambda_{P}{\lambda}_{V}}{m_{V}\big((p-q)^{2}-m_{V}^{2}\big)\big(q^{2}-m^{2}_{P}\big)}
+\int_{\Sigma}\frac{\rho^{h}(s_1,s_2)ds_1ds_2}{\big(s_1-q^2\big)\big(s_2-(p-q)^2\big)}+
{subtractions}.
\end{eqnarray}
The first term is the contribution of the ground-state and contains
the $g_{V P \gamma}$ coupling, while the hadronic spectral function
$\rho^h(s_1,s_2)$ represents the contribution of higher resonances
and continuum states. The integration region in the $(s_1,s_2)$ -
plane is denoted by $\Sigma$ and one may take $(s_1)^a + (s_2)^a
\leq (s_0)^a$, where $s_0$ is the effective threshold in the double
dispersion relation. It is relevant with $s^{V}_0$ and $s^P_0$,
which are the effective thresholds in the vector meson and the
pseudoscalar meson channels, respectively. How to take their values
will be discussed in the next section. The subtraction terms isn't
considered in the sum rules, because they will be removed by a
double Borel transformation.

Next, we calculate the correlation function from QCD side by using
light cone operator product expansion method, in which we work with
large momenta, i.e., $-q^2$ and $-(p - q)^2$ are both large. The
correlation function, then, can be calculated as an expansion near
to the light cone $x^2\approx0$. The expansion involves matrix
elements of the nonlocal operators between vacuum and the photon
states in terms of the photon DAs with increasing twist. At the same
time, the full quark propagator of the light quark
\cite{propagator1,propagator2} in the presents of gluonic and
electromagnetic background fields is used in this calculation, and
it is given as
\begin{eqnarray}
i S(x,0)&=&<0|T\{\bar{q}(x)q(0)\}|0>
=i\frac{\slashed{x}}{2\pi^2x^4}-\frac{m_q}{4\pi^2x^2}-\frac{<\bar{q}q>}{12}(1+i\frac{m_q\slashed{x}}{4})
-\frac{x^2}{192}m^2_0<\bar{q}q>(1+i\frac{m_q\slashed{x}}{6})
\nonumber\\
& &-i g_s \int^1_0 d
v\{\frac{\slashed{x}}{16\pi^2x^2}(\bar{\nu}\slashed{x}\sigma_{\mu\nu}+\nu\sigma_{\mu\nu}\slashed{x})G^{\mu\nu}(vx)
-i\frac{m_q}{32\pi^2}G^{\mu\nu}(vx)\sigma_{\mu\nu}\ln(\frac{-x^2\Lambda^2}{4}+2\gamma_E)\}
\nonumber\\
& &-i e_q \int^1_0 d
v\{\frac{\slashed{x}}{16\pi^2x^2}(\bar{\nu}\slashed{x}\sigma_{\mu\nu}+\nu\sigma_{\mu\nu}\slashed{x})F^{\mu\nu}(vx)
-i\frac{m_q}{32\pi^2}F^{\mu\nu}(vx)\sigma_{\mu\nu}\ln(\frac{-x^2\Lambda^2}{4}+2\gamma_E)\}
+\cdots,
\end{eqnarray}
where $G_{\mu\nu}$ is the gluon field strength tensor and
$F_{\mu\nu}$ is the electromagnetic field strength tensor. Taking
into account the flavor SU(3)-breaking effect, we take the quark
mass as $m_u=m_d=0$ and $m_s\neq 0$ in our calculation.

We substitute the above propagator and the relevant photon DAs
\cite{photonDA1,photonDA2} into correlation function (1) and
integrate over the time-space coordinate x, we can obtain the
expression of the amplitude $T((p-q)^2,q^2)$ up to twist-4 accuracy.
The following taking place is a standard process in light-cone sum
rules. The hadronic spectral function $\rho^h(s_1,s_2)$, which was
used to control the contributions of excited states and of the
continuum, can be approximately evaluated by using quark-hadron
duality. To suppress the contributions of the excited and continuum
states and of the subtraction terms in the hadronic representation,
the double Borel transformation
\cite{doubBorel1,doubBorel2,doubBorel3} about the variables $-q^2$
and $-(p-q)^2$ was introduced and the useful formulas for
transformation as follow:
\begin{eqnarray}
\mathcal{B}_{M_1^2}\mathcal{B}_{M_2^2}\bigg\{\frac{\Gamma(\alpha)}{\big[-uq^2-\bar{u}(p-q)^2\big]^{\alpha}}\bigg\}=
\frac{(M^2)^{2-\alpha}}{M_1^2M_2^2}\delta(u-u_0)
\end{eqnarray}and
\begin{eqnarray}
\mathcal{B}_{M_1^2}\mathcal{B}_{M_2^2}\bigg\{\frac{1}{\big((p-q)^2-m_{V}^2\big)\big(q^2-m_P^2\big)}\bigg\}
=\frac{1}{M_1^2M_2^2}e^{-\frac{m_P^2}{M_1^2}}e^{-\frac{m_{V}^2}{M_2^2}}.
\end{eqnarray}
Here $M^2=\frac{M_1^2M_2^2}{M_1^2+M_2^2}$, $
u_0=\frac{M_1^2}{M_1^2+M_2^2}$, $M_1^2$ and $M_2^2$ are the Borel
parameters associated with $-q^2$ and $-(p-q)^2$, respectively.

After the above lengthy calculation, we obtained the final result
for the coupling $g_{V P\gamma}$:
\begin{eqnarray}
g_{VP\gamma}&=&\frac{1}{\lambda_V
\lambda_P}e^{\frac{m_V^2}{M_2^2}+\frac{m_P^2}{M_1^2}}\bigg(\frac{A_{VP}
<\bar{q}q>}{2}X+B_{VP}<\bar{s}s>(X+B)\bigg),
\end{eqnarray}
where $X=A+I_G(u_0)+I_F(u_0)$,
\begin{eqnarray}
A&=&-M^2\chi\bigg[\varphi_{\gamma}(u_0)+\sum_k
b_ke^{-\frac{s_0}{M^2}}(\frac{1}{2})^kk!\sum_{n=0}^{k}\frac{(1-2u_0)^{k-n}}{(k-n)!n!}
\sum^{k-n}_{j=0}\frac{(\frac{s_0}{M^2})^j}{j!}\bigg]+\frac{\mathcal{A}(u_0)}{4},
\end{eqnarray}
\begin{eqnarray}
B&=&-\frac{m_s}{\pi^2}M^2(1-e^{-\frac{s_0}{M^2}})(1-\frac{\gamma_E}{2})
\nonumber\\
&&-\frac{f_{3\gamma}}{2}m_s{\phi}^{(a)}(u_0)-\frac{2<\bar{s}s>}{3}+\frac{2<\bar{s}s>}{3M^2}m^2_s+\frac{<\bar{s}s>}{3{\pi}^2M^4}m^2_sm^2_0,
\end{eqnarray}
\begin{eqnarray}
I_G(u_0)&=&\int_0^{u_0}d\alpha_q\int_0^{1-u_0}d\alpha_{\bar{q}}\frac{1}{1-\alpha_q-\alpha_{\bar{q}}}
[\mathcal{T}_1(\alpha_{i})-\mathcal{T}_2(\alpha_{i})+\mathcal{T}_3(\alpha_{i})-\mathcal{T}_4(\alpha_{i})
-\mathcal{S}(\alpha_{i})
\nonumber\\
{}{}&&-\tilde{\mathcal{S}}(\alpha_i)\bigg)]
-2\int_0^{u_0}d\alpha_q\int_0^{1-u_0}d\alpha_{\bar{q}}\frac{1-u_0-\alpha_{\bar{q}}}{(1-\alpha_q
-\alpha_{\bar{q}})^2}[\mathcal{T}_3(\alpha_i)-\mathcal{T}_4(\alpha_i)-\tilde{\mathcal{S}}(\alpha_i)]
\end{eqnarray}
and
\begin{eqnarray}
I_F(u_0)=-\int_0^{u_0}d\alpha_q\int_0^{1-u_0}d\alpha_{\bar{q}}\frac{1}{1-\alpha_q-\alpha_{\bar{q}}}
[\mathcal{S}_{\gamma}(\alpha_i)+\mathcal{T}_4^{\gamma}(\alpha_i)]+2\int_0^{u_0}d\alpha_q\int_0^{1-u_0}
d\alpha_{\bar{q}}\frac{1-u_0-\alpha_{\bar{q}}}{(1-\alpha_q-\alpha_{\bar{q}})^2}\mathcal{T}_4^{\gamma}(\alpha_i).
\end{eqnarray}
Here $\chi$ is the magnetic susceptibility of the quark condensate,
which has been introduced in the pioneering work \cite{msp} for
proton and neutron magnetic moments. In our next numerical analysis,
we take $\chi(1 GeV^2)=-3.15\pm0.3GeV^{-2}$ which was obtained by
using QCD sum rules analysis of two-points correlation function
\cite{photonDA1}. $\varphi_{\gamma}(u)$ is the photon leading twist
distribution amplitude, $\phi^{(a)}(u_0)$ is the photon twist-3
distribution amplitude, $\mathcal{A}(u)$,
$\mathcal{T}_j(\alpha_i)(j=1,2,3,4)$,
$\mathcal{T}_4^{\gamma}(\alpha_i)$, $\mathcal{S}(\alpha_i)$,
$\tilde{\mathcal{S}}(\alpha_i)$, and
$\mathcal{S}_{\gamma}(\alpha_i)(i=q,\bar{q},g$ and
$\alpha_g=1-\alpha_q-\alpha_{\bar{q}})$ are the photon twist-4 DAs.
Their detailed expression are given in Appendix A.
$f_{3\gamma}=-(0.0039\pm0.0020)GeV^2$ \cite{photonDA1} is the
nonperturbative constant to describe the photon twist-3 DAs and
$\gamma_E$ is Euler constant. $b_k$ are the coefficients of the
leading twist-2 distribution amplitude $\varphi_{\gamma}(u)$ by
exploited as a power series in $(1-u)$, $\varphi_{\gamma}(u)=\sum_k
b_k (1-u)^k$. The coupling $g_{V P \gamma}$ for the $P\rightarrow
V\gamma$ decay can also be calculated by the similar approach. A
summary of the results is presented in Table 1.
%% table 1
\begin{table}[htbp!]\small\renewcommand{\addcontentsline}[3]{}
\begin{center}
\begin{tabular}{|c|c|c|c|c|}
\multicolumn{5}{l}{}\\
\hline
\multirow{2}[3]*{\backslashbox{$V$}{$P$}}&\multicolumn{2}{c|}{
$\eta$}&\multicolumn{2}{c|}{$\eta'$}\\
\cline{2-5}
&$A_{VP}$&$B_{VP}$&$A_{VP}$&$B_{VP}$\\
\hline
$\phi$&$m_\phi(e_u+e_d)\sin\beta\cos\varphi $&$-m_\phi e_s\cos\beta\sin\varphi $&$m_\phi(e_u+e_d)\sin\beta\sin\varphi $&$m_\phi e_s\cos\beta\cos\varphi $\\
\hline
$\omega$&$m_\omega(e_u+e_d)\cos\beta\cos\varphi $&$m_\omega e_s\sin\beta\sin\varphi $&$m_{\eta'}(e_u+e_d)\cos\beta\sin\varphi $&$-m_{\eta'} e_s\sin\beta\cos\varphi $\\
\hline
$\rho$&$m_\rho(e_u-e_d)\cos\varphi $&$0$&$m_{\eta'}(e_u-e_d)\sin\varphi $&$0$\\
\hline
\end{tabular}
\end{center}
\begin{center}
Table 1:The parameters for the coupling $g_{V P \gamma}$ in the
radiative decays of the light mesons.
\end{center}
\end{table}

\section{Numerical Analysis}

Now, we present our numerical analysis of the coupling constants
$g_{V P \gamma}$ for the $V\rightarrow P\gamma$ and $P\rightarrow V
\gamma$ decays. In order to obtain numerical results of the sum
rules from Eq.(13) and Table.1, we take the input parameters as
usual: $m_\phi=1.02 GeV$, $m_\omega=0.782 GeV$, $m_\rho=0.77 GeV$,
$m_\eta=0.55 GeV$, $m_{\eta^\prime}=0.958 GeV$, $m_s=0.156 GeV$,
$\langle \bar{q}q\rangle=-(0.24 GeV)^3$, $\langle
\bar{s}s\rangle=0.8 \langle \bar{q}q\rangle$. The values of overlap
amplitudes $\lambda_V$ from Eq.(5) are $\lambda_\phi=0.250\pm0.009
GeV^2$,$\lambda_\omega=0.162\pm0.004
GeV^2$,$\lambda_\rho=0.150\pm0.003 GeV^2$ and their numerical
analysis was presented in Appendix B. There haven't the numerical
results for the overlap amplitudes $\lambda_P$ from Eq.(6), but
their expressions with the mixing angle $\varphi$ were given by
using QCD sum rules in Appendix B.

The following issue is to discuss how to reasonably choose the
effective thresholds $s_0, s_0^V, s_0^P$ for the sum rules of the
coupling $g_{V P \gamma}$. $s_0^V$ and $s_0^P$ are the effective
thresholds in the vector meson and the pseudoscalar meson channels,
respectively. In general, their values ranges from the mass square
of the ground state to the mass square of the first exciting state.
According to data of PDG \cite{pdg}, we can find the first exciting
states of the above meson. They are $\phi(1680)$ for $\phi$ meson,
$\omega(1420)$ for $\omega$ meson, $\rho(1450)$ for $\rho$ meson,
$\eta(1295)$ for $\eta$ meson, $\eta^\prime(1405)$ for $\eta^\prime$
meson. While taking into account the especial property of the double
spectral function at $s_1=s_2$, we obtain the eventual range of the
effective threshold in the double dispersion relation:
$s_0\in(1.04,1.68)GeV^2$ for the coupling $g_{\phi \eta \gamma}$,
$s_0\in(1.04,1.97)GeV^2$ for the coupling $g_{\phi
\eta^\prime\gamma}$, $s_0\in(0.61,1.68)GeV^2$ for the coupling
$g_{\omega \eta \gamma}$, $s_0\in(0.59,1.68)GeV^2$ for the coupling
$g_{\rho \eta\gamma}$, $s_0\in(0.92,1.97)GeV^2$ for the coupling
$g_{\omega \eta^\prime\gamma}$ and the coupling $g_{\rho
\eta^\prime\gamma}$. In the next analysis, the effective threshold
$s_0$ will be strictly choose in the above ranges to fit the result
of the sum rules and experimental data.

In Fig.1, we discussed the Borel window of our sum rules. Here the
mixing angle $\varphi$ is fixed at $\varphi=40^\circ$, the Borel
parameter $M_2^2$ of every coupling constant takes a defined value
and the effective thresholds $s_0$ of the decay channel have
different value which belong to the above discussing ranges. We find
that there have a platform, where the coupling $g_{VP\gamma}$ is
practically independent of the Borel parameter $M_1^2$, $1.8
GeV^2\leq M_1^2 \leq 2.8 GeV^2$ for the $\eta$ channel and $1.5
GeV^2\leq M_1^2 \leq 2.5 GeV^2$ for the $\eta^\prime$ channel. So
our sum rules are reasonable and significative.

In Fig.2, we show theoretical and experimental values of the
couplings $g_{VP\gamma}$ as functions of the $\eta-\eta^\prime$
mixing angle $\varphi$ with the defined $s_0$, $M_2^2$ and the
variable $M_1^2$. The ranges of $M_1^2$ are the Borel windows from
Fig.1. So theoretical predictions are presented in the shadows in
Fig.2. By using the definition of $g_{VP\gamma}$ from Eq.(8), the
decay widths of $V\rightarrow P\gamma$ and $P\rightarrow V\gamma$
are written as
\begin{eqnarray}
\Gamma(V\rightarrow P\gamma)=\frac{\alpha g^2_{V
P\gamma}}{24}\frac{(m_V^2-m_P^2)^3}{m_V^5},~~ \Gamma(P\rightarrow
V\gamma)=\frac{\alpha g^2_{V
P\gamma}}{8}\frac{(m_P^2-m_V^2)^3}{m_P^5},
\end{eqnarray}
where $\alpha=1/137$ is the electromagnetic coupling constant.
Comparing Eq.(18) with experimental data \cite{pdg}, we obtain the
values of the coupling constants, which are presented in the
dot-dashed curves in Fig.2. Finally, the pseudoscalar mixing angle
$\varphi$ from different $V-\eta,\eta^\prime$ electromagnetic
coupling processes and their average value are list in Table.2.

%%ͼ1

\begin{figure}[htbp!]
\begin{center}
\setlength{\unitlength}{1cm}
\begin{center}
\begin{minipage}[t]{7.5cm}
\begin{picture}(8.5,5.5)(0,0.3)
\includegraphics*[scale=0.7,angle=0.]{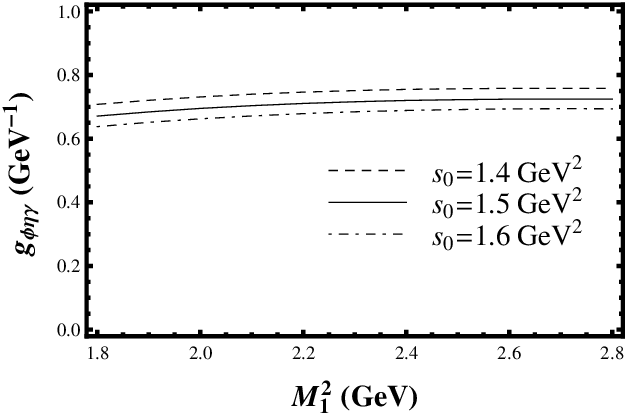}
\end{picture}\par
\begin{center}
(a)
\end{center}
\end{minipage}
%\hfill
\hspace{0.5cm}
\begin{minipage}[t]{7.5cm}
\begin{picture}(8.5,5.5)(0,0.3)
\includegraphics*[scale=0.7,angle=0.]{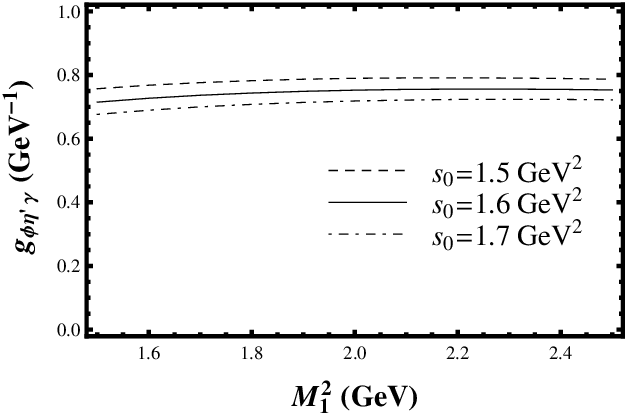}
\end{picture}\par
\begin{center}
(b)
\end{center}
\end{minipage}
\end{center}

\vspace{-1cm}
\begin{center}
\begin{minipage}[t]{7.5cm}
\begin{picture}(8.5,5.5)
\includegraphics*[scale=0.7,angle=0.]{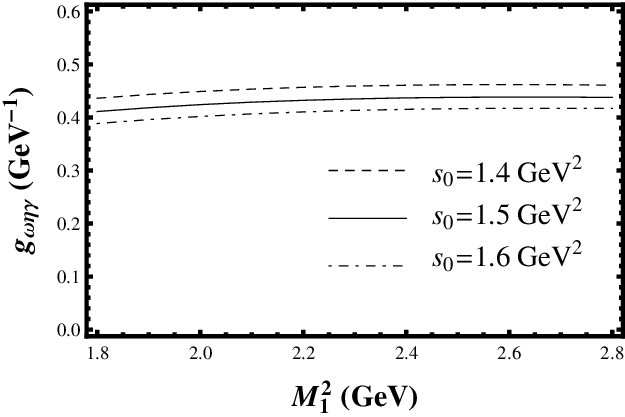}
\end{picture}\par
\begin{center}
(c)
\end{center}
\end{minipage}
%\hfill
\hspace{0.5cm}
\begin{minipage}[t]{7.5cm}
\begin{picture}(8.5,5.5)
\includegraphics*[scale=0.7,angle=0.]{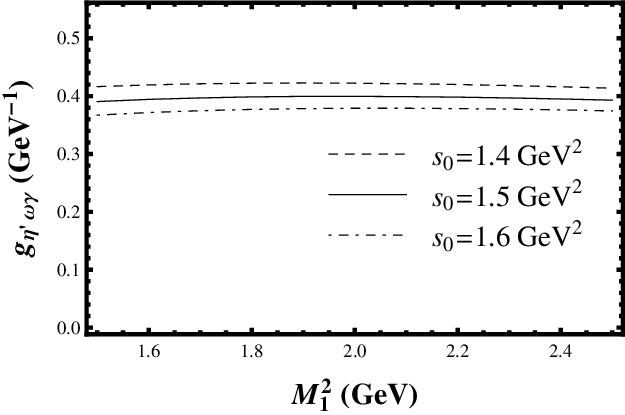}
\end{picture}\par
\begin{center}
(d)
\end{center}
\end{minipage}
\end{center}

\vspace{-1cm}
\begin{center}
\begin{minipage}[t]{7.5cm}
\begin{picture}(8.5,5.5)
\includegraphics*[scale=0.7,angle=0.]{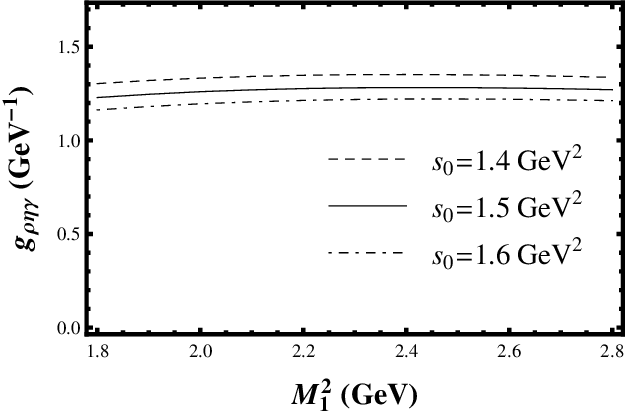}
\end{picture}\par
\begin{center}
(e)
\end{center}
\end{minipage}
%\hfill
\hspace{0.5cm}
\begin{minipage}[t]{7.5cm}
\begin{picture}(8.5,5.5)
\includegraphics*[scale=0.7,angle=0.]{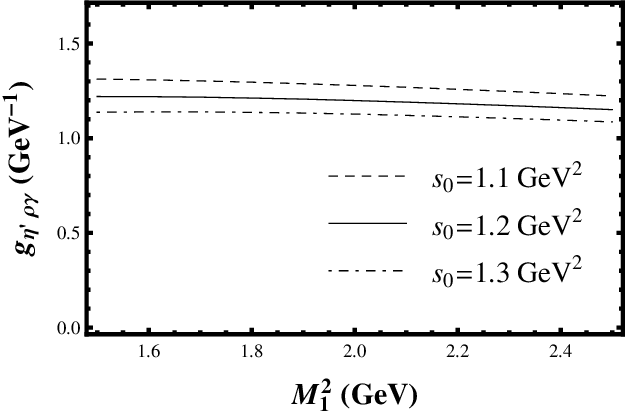}
\end{picture}\par
\begin{center}
(f)
\end{center}
\end{minipage}
\end{center}
\caption{The the coupling constants $g_{VP\gamma}$ and
$g_{PV\gamma}$ as a function of the Borel parameter $M_1^2$ for
different values of the threshold parameters $s_0$ with defined
$M_2^2$. (a) $M_2^2=2.0 GeV^2$; (b) $M_2^2=2.3 GeV^2$; (c)
$M_2^2=1.9 GeV^2$; (d) $M_2^2=2.1 GeV^2$; (e) $M_2^2=1.7 GeV^2$; (f)
$M_2^2=1.9GeV^2$.}
\end{center}
\end{figure}

%%ͼ2
\begin{figure}[htbp!]
\begin{center}
\setlength{\unitlength}{1cm}
\begin{center}
\begin{minipage}[t]{7.5cm}
\begin{picture}(8.5,5.5)(0,0.3)
\includegraphics*[scale=0.7,angle=0.]{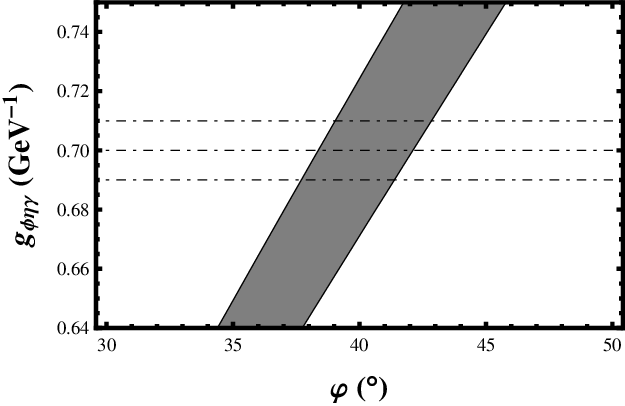}
\end{picture}\par
\begin{center}
(a)
\end{center}
\end{minipage}
%\hfill
\hspace{0.3cm}
\begin{minipage}[t]{7.5cm}
\begin{picture}(8.5,5.5)(0,0.3)
\includegraphics*[scale=0.7,angle=0.]{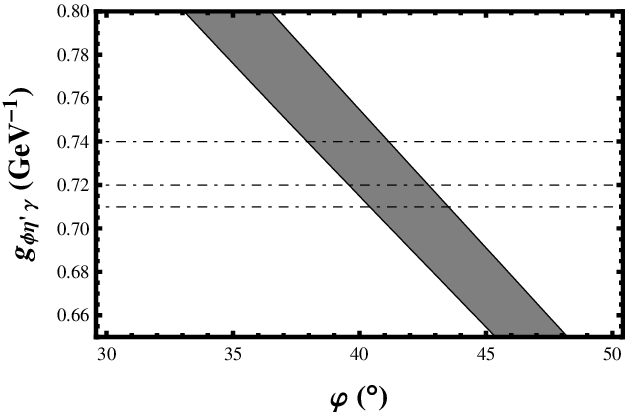}
\end{picture}\par
\begin{center}
(b)
\end{center}
\end{minipage}
\end{center}
\vspace{-1cm}

\begin{center}
\begin{minipage}[t]{7.5cm}
\begin{picture}(8.5,5.5)
\includegraphics*[scale=0.7,angle=0.]{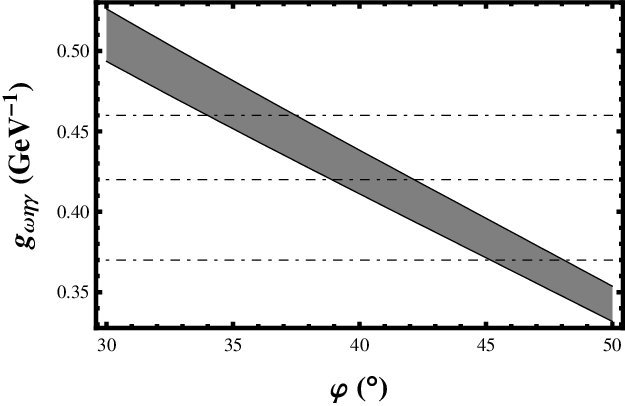}
\end{picture}\par
\begin{center}
(c)
\end{center}
\end{minipage}
%\hfill
\hspace{0.3cm}
\begin{minipage}[t]{7.5cm}
\begin{picture}(8.5,5.5)
\includegraphics*[scale=0.7,angle=0.]{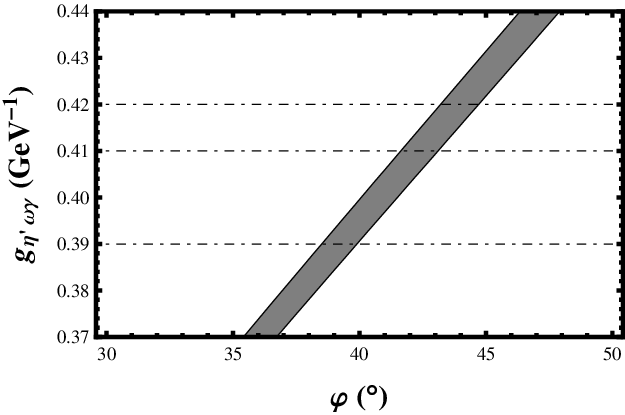}
\end{picture}\par
\begin{center}
(d)
\end{center}
\end{minipage}
\end{center}
\vspace{-1cm}
\begin{center}
\begin{minipage}[t]{7.5cm}
\begin{picture}(8.5,5.5)
\includegraphics*[scale=0.7,angle=0.]{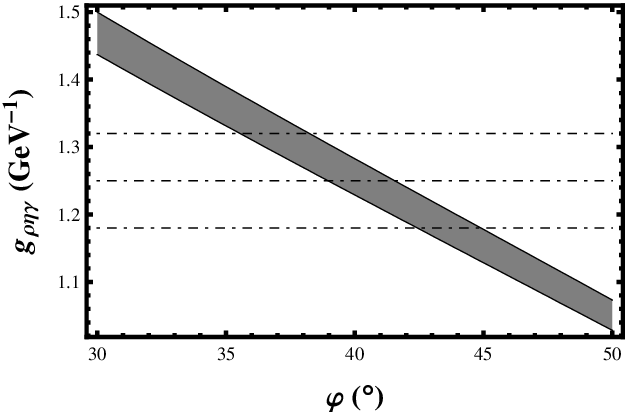}
\end{picture}\par
\begin{center}
(e)
\end{center}
\end{minipage}
%\hfill
\hspace{0.5cm}
\begin{minipage}[t]{7.5cm}
\begin{picture}(8.5,5.5)
\includegraphics*[scale=0.7,angle=0.]{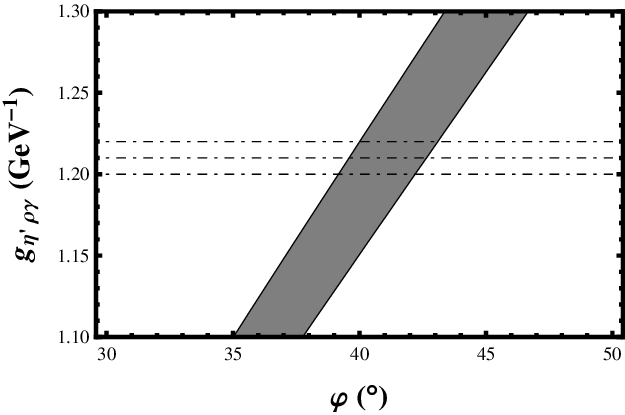}
\end{picture}\par
\begin{center}
(f)
\end{center}
\end{minipage}
\end{center}
\caption{Theoretical and experimental values of the couplings
$g_{VP\gamma}$ as functions of the $\eta-\eta^\prime$ mixing angle
$\varphi$. The shadows are theoretical predictions according to
Eq.(13) and the dot-dashed curves are the couplings extracted from
experimental data \cite{pdg} by Eq.(18). (a) $M_2^2=2.0 GeV^2$ and
$s_0=1.5 GeV^2$; (b) $M_2^2=2.3 GeV^2$ and $s_0=1.6 GeV^2$;
(c)$M_2^2=1.9 GeV^2$ and $s_0=1.5 GeV^2$; (d) $M_2^2=2.1 GeV^2$ and
$s_0=1.5 GeV^2$; (e) $M_2^2=1.7GeV^2$ and $s_0=1.5 GeV^2$; and (f)
$M_2^2=1.9GeV^2$ and $s_0=1.2 GeV^2$.}
\end{center}
\end{figure}

%%table 2
\begin{table}[htbp!]
\begin{center}
\begin{tabular}{|c|c|c|c|c|c|c|c|}
\hline $V(P)\rightarrow
P(V)\gamma$&$\phi\rightarrow\eta\gamma$&$\omega\rightarrow\eta\gamma$&$\rho\rightarrow\eta\gamma$&
$\phi\rightarrow\eta'\gamma$&$\eta'\rightarrow\omega\gamma$&$\eta'\rightarrow\rho\gamma$&$\varphi_{av}(^\circ)$\\
\hline
$\varphi(^\circ)$&$40.2\pm0.7$&$41.2\pm5.4$&$40.3\pm3.4$&$40.8\pm1.2$&$41.6\pm2.4$&$41.1\pm0.4$ &$40.9\pm0.5$\\
\hline
\end{tabular}
\end{center}
\begin{center}
{Table 2:The mixing angle $\varphi$ from different
$V-\eta,\eta^\prime$ electromagnetic coupling processes.}
\end{center}
\end{table}

\section{Conclusions}

We calculated the coupling $g_{V P \gamma}$ $(V=\phi,\omega,\rho$
and $P=\eta,\eta^\prime )$ of the $V\rightarrow P\gamma$ and
$P\rightarrow V\gamma$ electromagnetic decays in the light-cone QCD
sum rules. Comparing theoretical results and experimental data, we
extracted a new pseudoscalar mixing angle $\varphi=40.9\pm0.5^\circ$
in the quark-flavor basis. This result is in agreement with
Ref.\cite{qfs}, where the average $\varphi=39.3\pm1.0^\circ$.
Recently, the KLOE Collaboration \cite{ex1,ex2} has measured the
ratio
$R_\phi=\mathcal{B}({\phi\rightarrow\eta'\gamma})/\mathcal{B}({\phi\rightarrow\eta\gamma})=4.77\times
10^{-3}$, the pseudoscalar mixing angle $\varphi=41.4\pm1.6^\circ$
with the zero gluonium content for $\eta^\prime$ and
$\varphi=39.7\pm0.7^\circ$ with the gluonium content for
$\eta^\prime$. There is a little discrepancy between our theoretical
results, $R_\phi=4.85\times 10^{-3}$ and $\varphi=40.5\pm1.0^\circ$
from the $\phi\rightarrow \eta\gamma$ and $\phi\rightarrow
\eta^\prime\gamma$ decays, and the experimental results from KLOE. A
possible reason is that we should consider the contribution from the
gluonium content of $\eta^\prime$ meson in our calculation. This
will be our next work.

\begin{appendix}
\section*{Appendix A: The photon distribution amplitudes }

In this section, the clear expressions for the photon distribution
amplitudes are showed as\cite{photonDA1,photonDA2}
\begin{eqnarray*}
\varphi_{\gamma}(u)&=&6u\bar{u}\big(1+\varphi_2\mathcal{C}_2^{\frac{3}{2}}(u-\bar{u})\big),
\nonumber\\
\phi^{(a)}(u)&=&\big(1-(2u-1)^2\big)\big(5(2u-1)^2-1\big)\frac{5}{2}(1+\frac{9}{16}\omega^V_\gamma-\frac{3}{16}\omega^A_\gamma),
\nonumber\\
\mathcal{A}(u)&=&40u^2\bar{u}^2(3\kappa-\kappa^++1)+8(\zeta^+_2-3\zeta_2)\big[u\bar{u}(2+13u\bar{u})
\nonumber\\
&&{}+2u^3(10-15u+6u^2)ln(u)+2\bar{u}^3(10-15\bar{u}+6\bar{u}^2)ln(\bar{u})\big],
\nonumber\\
\mathcal{T}_1(\alpha_i)&=&-120(3\zeta_2+\zeta_2^+)(\alpha_{\bar{q}}-\alpha_q)\alpha_{\bar{q}}\alpha_q\alpha_g,
\nonumber\\
\mathcal{T}_2(\alpha_i)&=&30\alpha_g^2(\alpha_{\bar{q}}-\alpha_q)\big((\kappa-\kappa^+)
+(\zeta_1-\zeta_1^+)(1-2\alpha_g)+\zeta_2(3-4\alpha_g)\big),
\nonumber\\
\mathcal{T}_3(\alpha_i)&=&-120(3\zeta_2-\zeta_2^+)(\alpha_{\bar{q}}-\alpha_q)\alpha_{\bar{q}}\alpha_q\alpha_g,
\nonumber\\
\mathcal{T}_4(\alpha_i)&=&30\alpha_g^2(\alpha_{\bar{q}}-\alpha_q)\big((\kappa+\kappa^+)
+(\zeta_1+\zeta_1^+)(1-2\alpha_g)+\zeta_2(3-4\alpha_g)\big),
\nonumber\\
\mathcal{S}(\alpha_i)&=&30\alpha_g^2\{(\kappa+\kappa^+)(1-\alpha_g)+(\zeta_1+\zeta_1^+)
(1-\alpha_g)(1-2\alpha_g)
\nonumber\\
&&{}+\zeta_2[3(\alpha_{\bar{q}}-\alpha_q)^2-\alpha_g(1-\alpha_g)]\},
\nonumber\\
\tilde{\mathcal{S}}(\alpha_i)&=&-30\alpha_g^2\{(\kappa-\kappa^+)(1-\alpha_g)+(\zeta_1-\zeta_1^+)
(1-\alpha_g)(1-2\alpha_g)
\nonumber\\
&&{}+\zeta_2[3(\alpha_{\bar{q}}-\alpha_q)^2-\alpha_g(1-\alpha_g)]\},
\nonumber\\
\mathcal{S}_{\gamma}(\alpha_i)&=&60\alpha_g^2(\alpha_q+\alpha_{\bar{q}})(4-7(\alpha_{\bar{q}}+\alpha_q)),
\nonumber\\
\mathcal{T}_4^{\gamma}(\alpha_i)&=&60\alpha_g^2(\alpha_q-\alpha_{\bar{q}})(4-7(\alpha_{\bar{q}}+\alpha_q)).
\end{eqnarray*}
 Here $\varphi_{\gamma}(u)$ is the photon leading twist
distribution amplitude, $\phi^{(a)}(u_0)$ is the photon twist-3 DA,
$\mathcal{A}(u)$, $\mathcal{T}_j(\alpha_i)(j=1,2,3,4)$,
$\mathcal{T}_4^{\gamma}(\alpha_i)$, $\mathcal{S}(\alpha_i)$,
$\tilde{\mathcal{S}}(\alpha_i)$, and
$\mathcal{S}_{\gamma}(\alpha_i)(i=q,\bar{q},g$ and
$\alpha_g=1-\alpha_q-\alpha_{\bar{q}})$ are the photon twist 4 DAs.
The parameters appearing in the above DAs are given as
$\varphi_2=0$, $\kappa=0.2$, $\kappa^+=0$, $\zeta_1=0.4$,
$\omega^V_\gamma=3.8$, $\omega^A_\gamma=-2.1$, $\zeta_2=0.3$,
$\zeta_1^+=0$ and $\zeta_2^+=0$ at the scale $\mu=1 GeV$.

\section*{Appendix B: Calculation of Overlap Amplitude }
In this section, we present the discussion of the overlap amplitudes
for the vector mesons and pseudoscalar mesons. They are important
input parameters for the coupling $g_{VP\gamma}$ in our sum rules.
The overlap amplitudes $\lambda_{\eta}$, $\lambda_{\eta'}$ for
pseudoscalar meson $\eta$ and $\eta^\prime$ are given as\cite{lap}
\begin{equation*}
\lambda_\eta=\lambda_\eta^q \cos\varphi-\lambda_\eta^s \sin
\varphi,~~~~\lambda_{\eta^\prime}=\lambda_{\eta^\prime}^q
\cos\varphi+\lambda_{\eta^\prime}^s \sin \varphi
\end{equation*}
in QCD sum rules, where
\begin{eqnarray*}
(\lambda_{P}^q)^{2}&=&{e^\frac{m_P^2}{M^2}M^4}\{\frac{3}{8\pi^2M^2}[1-(1+\frac{s_0^P}{M^2})e^\frac{-s_0^P}{M^2}]
+\frac{\langle\frac{\alpha_s}{\pi}G^2\rangle}{8M^4}-m_q\langle\overline{q}q\rangle)\frac{1}{M^4}
+\frac{112\pi}{27M^6}<\sqrt{\alpha_s}\bar{q}q>^2\},\\
(\lambda_{P}^s)^{2}&=&{e^\frac{m_P^2}{M^2}M^4}\{\frac{3}{8\pi^2M^2}[1-(1+\frac{s_0^P}{M^2})e^\frac{-s_0^P}{M^2}]
+\frac{\langle\frac{\alpha_s}{\pi}G^2\rangle}{8M^4}-m_s\langle\overline{s}s\rangle)\frac{1}{M^4}
+\frac{112\pi}{27M^6}<\sqrt{\alpha_s}\bar{s}s>^2\}.
\end{eqnarray*}
Here $s_0^P$ is the effective threshold and the pseudoscalar mixing
angle $\varphi$ will be determined by numerical analysis of
$g_{VP\gamma}$.

With similar processes mentioned in the overlap amplitudes
$\lambda_P$ of pseudoscalar meson, the overlap amplitudes
$\lambda_V$ of vector mesons can be obtained and $\lambda_V^{q,s}$
are written as \cite{doubBorel3}:
\begin{eqnarray*}
(\lambda_V^q)^2&=&m_V^2M^2e^{m_{V}^2/M^2}\bigg[\frac{1}{4\pi^2}(1-e^{-s_0^{V}/M^2})
(1+\frac{\alpha_s}{\pi})+\frac{(m_u+m_d)}{2M^2}<\bar{u}u>
\nonumber\\
&+&\frac{<\frac{\alpha_s}{\pi}G_{\mu\nu}^aG^{a\mu\nu}>}{12M^4}-\frac{112\pi}{81}
\frac{\alpha_s<\bar{u}u>^2}{M^6}+\frac{m_u^3+m_d^3}{36M^8}<g_s\bar{u}\sigma_{\mu\nu}\frac{\lambda_a}{2}G^{a\mu\nu}u>\bigg],
\nonumber\\
(\lambda_V^s)^2&=&m_V^2M^2e^{m_{V}^2/M^2}\bigg[\frac{1}{4\pi^2}(1-e^{-s_0^{V}/M^2})
(1+\frac{\alpha_s}{\pi})+\frac{m_s}{M^2}<\bar{s}s>
\nonumber\\
&+&\frac{<\frac{\alpha_s}{\pi}G_{\mu\nu}^aG^{a\mu\nu}>}{12M^4}-\frac{112\pi}{81}
\frac{\alpha_s<\bar{s}s>^2}{M^6}+\frac{m_s^3}{18M^8}<g_s\bar{s}\sigma_{\mu\nu}\frac{\lambda_a}{2}G^{a\mu\nu}s>\bigg].
\end{eqnarray*}
And then we take the input parameters: $m_\phi=1.02 GeV$,
$m_\omega=0.782 GeV$, $m_\rho=0.77 GeV$, $m_u=0.005 GeV$, $m_d=0.008
GeV$, $m_s=0.156 GeV$, $\langle \bar{q}q\rangle=-(0.24 GeV)^3$,
$\langle \bar{s}s\rangle=0.8 \langle \bar{q}q\rangle$,
$\alpha_s=0.5$, $m_0^2=0.8GeV^2$,
$\frac{\alpha_s}{\pi}G^2=0.012GeV^4$,
$<g_s\bar{s}\sigma_{\mu\nu}\frac{\lambda_a}{2}G^{a\mu\nu}s>=m_0^2<\bar{s}s>$
into the above expressions and yield:
\begin{equation*}
\lambda_{\phi}=\lambda_\phi^q \sin\beta+\lambda_\phi^s\cos\beta
=0.250\pm0.009GeV^2,
\end{equation*}
\begin{equation*}
\lambda_{\omega}=\lambda_\omega^q\cos\beta
-\lambda_\omega^s\sin\beta =0.162\pm0.004GeV^2,
\end{equation*}
\begin{equation*}
\lambda_{\rho}= \lambda_\rho^q=0.150\pm0.003GeV^2.
\end{equation*}

\end{appendix}

\begin{center}
\section*{Acknowledgements}
\end{center}

We gratefully thanks to Profs. X. Q. Yu for a very useful
discussion. This work was supported by the Natural Science
Foundation of China, Grant Number 11005087, 10575083 and the
Fundamental Research Funds for the Central Universities, Grant
Number XDJK2009C185.

%\centerline{\rule{80mm}{0.1pt}} \vspace{2mm} \small
%\newpage

\end{document}